\input{aipcheck}

\documentclass[
    ,final            
  ]
  {aipproc}

\layoutstyle{6x9}

\def\barD{\overline D{}^0}

\def\barB{\overline B{}^0}
\def\Hbar{\overline{H}}

\def\beq{\begin{equation}}
\def\eeq{\end{equation}}
\def\bea{\begin{eqnarray}}
\def\eea{\end{eqnarray}}

\newcommand{\bra}[1]{\langle #1|}
\newcommand{\ket}[1]{|#1\rangle}

\newcommand{\bld}[1]{\mbox{\boldmath$#1$}}

\begin{document}

\title{\bld{$X(3872)$} in effective field theory}



\author{Alexey A Petrov}{
  address={Department of Physics and Astronomy, Wayne State University,
Detroit, Michigan 48201}
}



\begin{abstract}
We consider the implications from the possibility that the recently observed state
$X(3872)$ is a meson-antimeson molecule. We write an effective Lagrangian
consistent with the heavy-quark and chiral symmetries needed to describe $X(3872)$
and study its properties.
\end{abstract}

\maketitle


The unusual properties of $X(3872)$ state, recently discovered in the 
decay $X(3872) \to J/\psi \pi^+\pi^-$,
invited some speculations regarding its possible non-$c\bar c$ 
nature~\cite{Reviews,MolExp}. Since its mass lies
tantalizingly close to the $D^{*0}\barD$ threshold of 3871.3~MeV, it is
tempting to interpret $X(3872)$ as a $D^{*0}\barD$ 
molecule with $J^{PC}=1^{++}$ quantum numbers~\cite{MoleculeX,Braaten}. 
Such molecular states can be studied using techniques of effective field 
theories (EFT).

This study is possible due to the multitude of scales present in QCD.
The extreme smallness of the binding energy,
$E_b=(m^{\phantom{l}}_{D^0}+m^{\phantom{l}}_{D^{0*}})-M^{\phantom{l}}_X=
-0.6 \pm 1.1~\mbox{MeV}$,
suggests that this state can play the role of the ``deuteron''~\cite{MoleculeX}
in meson-meson interactions. This fact allows us to use methods similar to 
those developed for the description of the deuteron, with the added benefit of 
heavy-quark symmetry. The tiny binding energy of this molecular state introduces
an energy scale which is much smaller than the mass of the lightest
particle, the pion, whose exchange can provide binding.
Then, a suitable effective Lagrangian describing such a system contains only 
heavy-meson degrees of freedom with interactions approximated by local 
four-boson terms constrained only by the symmetries of the theory. 
This approach is similar to the Weinberg's EFT description of the 
deuteron~\cite{Weinberg}. While its predictive power is somewhat limited,
several model-independent statements can be made. For instance, possible existence 
of a molecular state in $D^{*0}\barD$ channel does not imply a molecular 
state in the $D^{*0}\overline {D^*}^0$ or $D^{0}\barD$ channels.

The general effective Lagrangian consistent with heavy-quark spin and chiral symmetries
can be written as~\cite{AlFiky:2005jd}
\beq\label{Lagr}
{\cal L}={\cal L}_2+{\cal L}_4,
\eeq
where the two-body piece that describes the strong interactions of
the heavy mesons $P$ and $P^*$ ($P=B,D$) containing one heavy quark $Q$
is well known~\cite{Grinstein:1992qt}:
\bea\label{Lagr2}
{\cal L}_2 =-i \mbox{Tr} \left[ \Hbar^{(Q)} v \cdot D H^{(Q)} \right]
- \frac{1}{2 m^{\phantom{l}}_P} \mbox{Tr} \left[ \Hbar^{(Q)} D^2 H^{(Q)} \right]
+ ~\frac{\lambda_2}{m^{\phantom{l}}_P}  \mbox{Tr}
\left[ \Hbar^{(Q)} \sigma^{\mu\nu } H^{(Q)} \sigma_{\mu\nu} \right] + ...
\eea
where the ellipsis denotes terms with more derivatives or including explicit
factors of light quark masses, or describing pion-$H$ interactions
and antimeson degrees of freedom $H_a^{(\overline{Q})}$ and $H_a^{(\overline{Q})\dagger}$.
A superfield describing the doublet of pseudoscalar heavy-meson 
fields $P_a = \left(P^0, P^+\right)$ and their vector counterparts with 
$v\cdot P^{*(Q)}_{a}=0$, is defined as 
$H_a^{(Q)}=(1+\not{v})\left[P^{*(Q)}_{a\mu} \gamma^\mu - P_a^{(Q)} \gamma_5
\right]/2$ (see~\cite{Grinstein:1992qt}). The third term in Eq.~(\ref{Lagr2}) accounts 
for the $P-P^*$ mass difference $\Delta\equiv m^{\phantom{l}}_{P^*}-m^{\phantom{l}}_P=
-2\lambda_2/m^{\phantom{l}}_P$. The four-body piece is~\cite{AlFiky:2005jd}
\bea\label{Lagr4}
{\cal L}_4=&-&\frac{C_1}{4} \mbox{Tr} \left[ \Hbar^{(Q)} H^{(Q)} \gamma_\mu \right]
\mbox{Tr} \left[ H^{(\overline{Q})} \Hbar^{(\overline{Q})} \gamma^\mu \right]
\nonumber \\
&-&\frac{C_2}{4} \mbox{Tr} \left[ \Hbar^{(Q)}  H^{(Q)} \gamma_\mu \gamma_5 \right]
\mbox{Tr} \left[ H^{(\overline{Q})} \Hbar^{(\overline{Q})} \gamma^\mu \gamma_5 \right].
\eea
Heavy-quark spin symmetry implies that the same Lagrangian governs the four-boson
interactions of {\it all} $P_a^{(*)}=D^{(*)}$ states. Indeed, not all of these states 
are bound. Here we shall concentrate on $X(3872)$, which we assume to be a bound state of two 
{\it neutral} bosons, $P_a\equiv P^0\equiv D$~\cite{MoleculeX}. Evaluating the traces 
yields for the $D\overline{D^*}$ sector
\bea\label{LocalLagr}
{\cal L}_{4,DD^*} = &-& C_1 D^{(c)\dagger} D^{(c)}
D^{*(\overline{c})\dagger}_\mu D^{* (\overline{c}) \mu}
- C_1 D^{*(c)\dagger}_\mu D^{*(c) \mu}
D^{(\overline{c})\dagger} D^{(\overline{c})} \nonumber \\
&+& C_2 D^{(c)\dagger} D^{*(c)}_\mu
D^{* (\overline{c})\dagger \mu} D^{(\overline{c})}
+ C_2 D^{* (c)\dagger}_\mu D^{(c)}
D^{(\overline{c})\dagger} D^{* (\overline{c}) \mu}
+\dots
\eea
As we show later, the resulting binding energy depends on 
a {\it linear combination} of $C_1$ and $C_2$. Similarly, one obtains the 
component Lagrangian governing the interactions of $D$ and $\overline D$,
\beq\label{LocalLagrPP}
{\cal L}_{4,DD} = C_1 D^{(c)\dagger} D^{(c)}
D^{(\overline{c})\dagger} D^{(\overline{c})}.
\eeq
Clearly, one cannot relate the existence of the bound state in the
$D\overline{D^*}$ and $D\overline{D}$ channels, as the properties of
the latter will depend only on $C_1$.

The lowest-energy bound state of $D$ and $\overline{D^*}$ is an eigenstate of 
charge conjugation,
\beq\label{Eigenstate}
\ket{X_{\pm}}=\frac{1}{\sqrt{2}}\left[
\ket{D^* \overline{D}} \pm \ket{D \overline{D}^*}
\right].
\eeq
To find the bound-state energy of $X(3872)$ with $J^{PC}=1^{++}$,
we shall look for a pole of the transition amplitude $T_{++}=\bra{X_+}T\ket{X_+}$.
Defining $DD^*$-$DD^*$ transition amplitudes,
\bea\label{Ts}
T_{11}&=&\langle D^* \overline{D}| T | D^* \overline{D} \rangle, \quad
T_{12}=\langle D^* \overline{D}| T | D \overline{D}^* \rangle,
\nonumber \\
T_{21}&=&\langle D \overline{D}^*| T | D^* \overline{D} \rangle, \quad
T_{22}=\langle D \overline{D}^*| T | D \overline{D}^* \rangle,
\eea
we also have to include a ``bubble'' resummation of loop contributions, as existence 
of a bound state is related to a breakdown of perturbative expansion~\cite{Weinberg}. 
These amplitudes satisfy a system of Lippmann-Schwinger equations~\cite{AlFiky:2005jd}.
In an algebraic matrix form,
\bea\label{LSEMatrix}
\left(
\begin{array}{c}
T_{11} \\
T_{12} \\
T_{21} \\
T_{22}
\end{array}
\right)
=
\left(
\begin{array}{c}
-C_1 \\
C_2 \\
C_2 \\
-C_1
\end{array}
\right)+
i\widetilde{A} \left(
\begin{array}{cccc}
-C_1 & C_2 & 0 & 0 \\
C_2 & -C_1 & 0 & 0 \\
0 & 0 & -C_1 & C_2 \\
0 & 0 & C_2 & -C_1
\end{array}
\right)
\left(
\begin{array}{c}
T_{11} \\
T_{12} \\
T_{21} \\
T_{22}
\end{array}
\right).
\eea
The solution of Eq.~(\ref{LSEMatrix}) produces the $T_{++}$ amplitude,
\beq\label{Solution}
T_{++}=\frac{1}{2}\left( T_{11}+T_{12}+T_{21}+T_{22} \right)=
\frac{\lambda}{1-i\lambda \widetilde{A}},
\eeq
where $\lambda=C_2-C_1$ and $\widetilde{A}$ is a (divergent) integral
\bea\label{Integral}
\widetilde{A}= \frac{i}{4} 2 \mu^{\phantom{l}}_{DD^*}
\int \frac{d^3q}{(2\pi)^3} \frac{1}{\vec{q}^{\;2}-2\mu^{\phantom{l}}_{DD^*}\left(E-\Delta\right)
-i\epsilon} = -\frac{1}{8 \pi} \mu^{\phantom{l}}_{DD^*}
|\vec{p}| \sqrt{1-\frac{2 \mu^{\phantom{l}}_{DD^*}\Delta}{\vec{p}^{\;2}}}.
\eea
Here $E=\vec{p}^2/2\mu^{\phantom{l}}_{DD^*}$, $\mu^{\phantom{l}}_{DD^*}$ is the reduced mass of
the $DD^*$ system. The divergence of the integral of Eq.~(\ref{Integral}) is removed 
by renormalization. We chose to define a renormalized $\lambda^{\phantom{l}}_R$ within the $MS$
subtraction scheme in dimensional regularization, which does not introduce any
new dimensionfull scales into the problem. In this scheme the integral $\widetilde{A}$ is
finite, which corresponds to an implicit subtraction of power divergences in
Eq.~(\ref{Integral}). This implies for the transition amplitude
\bea\label{FinAmp}
T_{++}=
\frac{\lambda^{\phantom{l}}_R}{1+(i/{8\pi})\lambda^{\phantom{l}}_R\, \mu^{\phantom{l}}_{DD^*}
|\vec{p}|
\sqrt{1-2 \mu^{\phantom{l}}_{DD^*}\Delta/{\vec{p}^{\;2}}}}.
\eea
The position of the pole of the molecular state on the energy scale
can be read off Eq.~(\ref{FinAmp}),
\beq\label{Pole}
E_{\rm Pole}=\frac{32 \pi^2}{\lambda_R^2 \mu_{DD^*}^3}-\Delta.
\eeq
Recalling the definition of binding energy $E_b$ and that 
$m^{\phantom{l}}_{D^*}$ = $m^{\phantom{l}}_{D}$ +  $\Delta$, we infer
\beq\label{Binding}
E_b=\frac{32 \pi^2}{\lambda_R^2 \mu_{DD^*}^3}.
\eeq
Assuming $E_b$ = 0.5 MeV, which is one sigma below the central value~\cite{MolExp}, 
and the experimental values for the masses~\cite{PDG}, we obtain 
$\lambda^{\phantom{l}}_R \simeq 8.4 \times 10^{-4} \ {\rm MeV}^{-2}$. 

Similar considerations apply to $D^0 \barD$ state, in which case the starting 
point is the Lagrangian term in Eq.~(\ref{LocalLagrPP}). Since it involves only a
single term, the calculations are actually easier and involve only
one Lippmann-Schwinger equation. The resulting binding energy is then~\cite{AlFiky:2005jd}
\beq\label{BindingC}
E_b=\frac{256 \pi^2}{C_{1R}^2 m_D^3}.
\eeq
Examining Eq.~(\ref{BindingC}) we immediately notice that the existence of
a bound state in the $D^*\overline{D}$ channel does not dictate the properties
of a possible bound state in the $D^0 \barD$ or $B^0 \barB$ channels, since $C_1$ and
$C_2$ are generally not related to each other.

We have used an effective field theory approach in the analysis of the
likely molecular state $X(3872)$, by describing its binding
interaction with contact terms in a heavy-quark symmetric
Lagrangian. The flexibility of this description allows us to ignore the
details of the interaction and to concentrate on its effects, namely a
shallow bound state and a large scattering length. Other applications are
possible, such as application to phenomenology~\cite{ExpX}. One can also show that 
if $X(3872)$ is indeed a molecular bound state of $D^{*0}$ and $\barD$ mesons, 
heavy-quark power counting implies the existence of the molecular bound state $X_b$ 
of $B^{*0}$ and $\barB$ with the mass of 10604 MeV~\cite{AlFiky:2005jd}. This work 
was supported in part by the U.S.\ National Science Foundation under Grant PHY--0244853, 
and by the U.S.\ Department of Energy under Contract DE-FG02-96ER41005.




\end{document}